\documentstyle[sprocl]{article}

\def\Journal#1#2#3#4{{#1} {\bf #2}, #3 (#4)}

\def\NPB{{\em Nucl. Phys.} B}
\def\PLB{{\em Phys. Lett.}  B}
\def\PRL{\em Phys. Rev. Lett.}
\def\PRD{{\em Phys. Rev.} D}

\def\PREP{{\em  Phys. Rept.}}

\def\mxth{\mathsurround=0pt }
\def\xversim#1#2{\lower2.pt\vbox{\baselineskip0pt \lineskip-.5pt
  \ialign{$\mxth#1\hfil##\hfil$\crcr#2\crcr\sim\crcr}}}             
                                    
\def\simle{\mathrel{\mathpalette\xversim <}}                                    
\newcommand{\mtt}{\mbox{$m_{3/2}$}} 
\newcommand{\mtts}{\mbox{$m^2_{3/2}$}} 

\newcommand{\mpl}{\mbox{$M_{pl}$}}  
 
\begin{document}

\title{$Z'$ PHYSICS FROM STRINGS \footnote{Invited talk
presented at PASCOS-98, Northeastern University, March 1998,
based on work done in collaboration 
with J. Cleaver, M. Cveti\v c, D. Demir,
J. R. Espinosa, L. Everett, and J. Wang.} }

\author{PAUL LANGACKER}

\address{Department of Physics and Astronomy \\ 
          University of Pennsylvania, Philadelphia PA 19104-6396, USA\\E-mail: 
pgl@langacker.hep.upenn.edu} 

\hfill UPR-0806T, hep-ph/9805486
\vspace{.5cm}

\maketitle\abstracts{ Many extensions of the standard model, especially
grand unified theories and superstring models, predict the existence of
additional $Z'$ bosons and associated exotic chiral supermultiplets. It 
has recently been argued that for classes of string motivated models with
supergravity mediated supersymmetry breaking there are two scenarios for the additional 
$Z'$s: either the mass is in the accessible range $< O$(1 TeV), providing a natural
solution to the $\mu$ problem and implications for the Higgs and sparticle masses and
for the LSP; or, when the breaking is associated with a $D$-flat direction, at an
intermediate scale, providing a possible explanation for the hierarchies of
quark and charged lepton masses and new possibilities for neutrino masses. Related work,
examining the detailed structure of specific perturbative string vacua for $D$ and
$F$-flat directions, surviving $U(1)$'s and exotics, and effective couplings, is 
briefly described. }

\newpage
\setcounter{page}{1}
\setcounter{footnote}{0}

\author{PAUL LANGACKER}

\address{Department of Physics and Astronomy \\ 
          University of Pennsylvania, Philadelphia PA 19104-6396, USA\\E-mail: 
pgl@langacker.hep.upenn.edu}

\title{$Z'$ PHYSICS FROM STRINGS \footnote{Work in collaboration 
with J. Cleaver, M. Cveti\v c, D. Demir,
J. R. Espinosa, L. Everett, and J. Wang.} }


\maketitle\abstracts{ Many extensions of the standard model, especially
grand unified theories and superstring models, predict the existence of
additional $Z'$ bosons and associated exotic chiral supermultiplets. It 
has recently been argued that for classes of string motivated models with
supergravity mediated supersymmetry breaking there are two scenarios for the additional 
$Z'$s: either the mass is in the accessible range $< O$(1 TeV), providing a natural
solution to the $\mu$ problem and implications for the Higgs and sparticle masses and
for the LSP; or, when the breaking is associated with a $D$-flat direction, at an
intermediate scale, providing a possible explanation for the hierarchies of
quark and charged lepton masses and new possibilities for neutrino masses. Related work,
examining the detailed structure of specific perturbative string vacua for $D$ and
$F$-flat directions, surviving $U(1)$'s and exotics, and effective couplings, is 
briefly described. }

\section{$Z'$ Phenomenology}
If the standard model (SM) gauge group is extended by an additional $U(1)$, then
the mass eigenstates $Z_{1,2}$ will be mixtures of the SM $Z$ and new $Z'$ with
mixing angle $\theta$. There are stringent limits on $M_{Z_2}$ and $\theta$
from precision $Z$ pole and neutral current experiments \cite{cl3}, 
because:
(i) $M_{Z_1}$ is shifted from the SM prediction by mixing; (ii) the $Z_1$ couplings
are changed by the mixing; (iii) $Z_2$ exchange may be important in neutral current
amplitudes. There are also Tevatron \cite{Tevatron}
limits on $M_{Z_2}$ from the non-observation of
$Z_2$ decays into $e^+ e^-$ or $\mu^+ \mu^-$. The limits are model dependent, depending
not only on the chiral couplings to $e, \nu, u$, and $d$, but (in the case of the
direct production limits) on the number of open decay channels into exotics, superpartners,
etc. Typically, for $Z'$ properties motivated by grand unification (GUTs) one
finds $M_{Z_2} >$ 600 -- 1000 GeV and $|\theta| <$ few $\times 10^{-3}$. For $M_{Z_2}
\gg M_{Z_1}$ one expects $\theta \sim C g_1' M_{Z_1}^2/G M_{Z_2}^2$,
where $G = \sqrt{g^2 + g_Y^2}$ and $g_1'$ are respectively
 the ordinary  and new $U(1)$ gauge couplings ($g_Y$ is the weak hypercharge
coupling), and $C$
depends on the Higgs charges under the
$U(1)'$ and their VEVs.
The most stringent limits on $M_{Z_2}$, which occur in those specific models in which $C$
is fixed,  actually come from $\theta$. 
For models with suppressed couplings to ordinary fermions 
\cite{suppressed}, such as leptophobic models,  much smaller $M_{Z_2}$
is allowed
 (e.g., 150 GeV; one could even have $M_{Z_2} < M_{Z_1}$, where $Z_1$
is the boson that is mainly the SM one), as is larger $|\theta| <$ few $\times 10^{-2}$.

It should be possible to extend the direct limits on $Z'$ with
GUT-type couplings to around a TeV at the Tevatron. At the LHC (with 100 fb$^{-1}$), one
should be able to discover a $Z'$ via its leptonic decays up to around
4 TeV \cite{cg,cl3}, well above the
range $M_{Z'} <$ 1 TeV expected in superstring theories \cite{cl12}.
At an NLC (500 GeV, 50 fb$^{-1}$) one  has  $e^+ e^- \rightarrow
\gamma, Z, Z' \rightarrow e^+ e^-, \mu^+ \mu^-, q \bar{q}, c\bar{c},
b\bar{b}$. Observations of cross sections, forward-backward and polarization
asymmetries, etc., should allow a sensitivity to a (virtual) $Z'$ up to
$\sim$ 1-3 TeV, increasing rapidly with energy \cite{cg}. Once a $Z'$ is
observed, one will want to determine not only its mass and mixing, but its
chiral couplings to identify its origin. At the LHC, a combination of
forward-backward asymmetries (as a function of rapidity), rapidity distributions,
rare decays ($Z' \rightarrow W \ell \nu$), and associated production of
$Z' Z, Z'W, Z' \gamma$ should provide significant diagnostic ability up
to 1 - 2 TeV \cite{cg}, with the information provided by the LHC and NLC
complementary.

\section{String Motivated Models}
\label{stringmotivated}
It is well known that that electroweak (EW) breaking in the MSSM with supergravity
mediated supersymmetry breaking (SUGRA) can
be radiative; i.e., a positive Higgs mass square from SUSY breaking at the Planck scale
can be driven negative at low energy due to the large Yukawa coupling associated
with the $t$ quark. In  perturbative
string models there are often extra non-anomalous $U(1)$'s which  are not broken at
the string scale. These can be broken radiatively \cite{cl12}, either at the
electroweak scale (i.e., less than 1 TeV) \cite{cdeel}, 
or, when the breaking is associated with a $D$-flat
direction, at an intermediate scale \cite{cceel1}.

\subsection{Electroweak Breaking}
In the ordinary MSSM the potential for the two Higgs doublets is
$V(H_1,H_2) = V_F + V_D + V_S$, where
\begin{eqnarray}
V_F & = & \mu^2 \left( |H_1|^2 + |H_2|^2 \right) \nonumber \\ 
V_D & = & \frac{G^2}{8} \left( |H_1|^2 - |H_2|^2 \right)^2 \nonumber \\ 
V_S & = & m_1^2 |H_1|^2 + m_2^2 |H_2|^2 - \left(B \mu H_1 \cdot H_2 + {\rm h.c.}
\right). 
\label{MSSMV}
\end{eqnarray}
(A term in $V_D$ involving charged fields has been omitted.)
The $F$ term $V_F$ is derived from the superpotential
\begin{equation}
W = \mu \hat{H}_1 \cdot \hat{H}_2 + h_Q \hat{u}^c_3 \hat{Q}_3 \cdot \hat{H}_2.
\label{MSSMW}
\end{equation}
Unlike the ordinary SM, in which the quartic coefficient $\lambda$ in the Higgs potential
is arbitrary, the coefficient $G^2 = g^2 + g_Y^2$ of the quartic $D$ term $V_D$ is associated
with gauge couplings, leading to the the upper bound on the lightest Higgs
scalar $m_{h_1^0} < M_Z$ (tree level) or $\simle$ 130 GeV (including loops).
The EW scale is $v^2 = v_1^2 + v_2^2 = (246 \ {\rm GeV})^2$, where $v_i = \sqrt{2}
\langle H^0_i \rangle$, and $M_Z = G v/2$. The scale of
$v$ is set not only  by  the
soft SUSY breaking parameters $m_i^2$ and $B$ in $V_S$, but also by the
supersymmetry preserving parameter $\mu$.

In SUGRA models one assumes that SUSY is broken in a hidden sector
at some intermediate scale
$M_I$ and then transmitted to the observable sector by supergravity. The soft breaking
parameters are then all of the same order of magnitude $ \mtt \sim$ 1 TeV (e.g., if
$\mtt \sim m_I^2/\mpl$, where \mpl \ is the Planck scale, then $M_I \sim 10^{11}$ GeV).
In particular, all  scalars in the theory  (Higgs, squarks, sleptons) typically
acquire positive mass squares of $O(\mtts)$ at \mpl. (Universal soft
breaking, which we do not assume, is the stronger assumption that the scalar
mass squares are all equal at \mpl.) This is the wrong sign for EW breaking.
However, for sufficiently large $h_Q = O(1)$ the Yukawa interactions can drive
$m_2^2$ negative (and of $O(-\mtts)$) at low energies. Hence, radiative breaking
requires a large $m_t$, consistent with the experimental value $\sim$ 175 GeV.

Thus, the SUGRA mechanism can yield the needed soft parameters. However,
one also requires $\mu = O(\mtt)$. Since $\mu$ is a supersymmetric parameter,
this requires fine-tuning in the context of the MSSM, the famous $\mu$ 
problem \cite{muprob}.
If one has some mechanism to force $\mu = 0$, then one can generate
an effective $\mu_{\rm eff} = O(\mtt)$ by several mechanisms, including:
(1) The Giudice-Masiero  mechanism \cite{giumas}, in which $\mu_{\rm eff}$ is transmitted
to the observable sector by SUGRA along with the soft breaking terms.
(2) The NMSSM \cite{NMSSM}, in which one introduces a SM singlet field $S$, with
superpotential terms $W_S = h_S \hat{S} \hat{H}_1 \cdot \hat{H}_2 +
\kappa \hat{S}^3$, so that  $\mu_{\rm eff} = h_S \langle S \rangle$. However,
the cubic term, needed to avoid an axion, allows a discrete symmetry and 
undesirable cosmological domain walls. (3) An extra gauge $U(1)'$ symmetry \cite{u1mu}
broken by the VEV of a SM singlet $S$ with  $W_S = h_S \hat{S} \hat{H}_1 \cdot \hat{H}_2$
can force $\mu = 0$ with $\mu_{\rm eff} = h_S \langle S \rangle$. Unlike the NMSSM
there is no domain wall problem. 

When the SUGRA MSSM is considered in the context of a class of perturbative string
models, one obtains in addition: (1) $\mu = 0$ by string selection rules.
(2) There are typically additional non-anomalous $U(1)$'s as well as exotic chiral
supermultiplets (which can play a role in radiative breaking). (3) The Yukawa couplings
at the string scale are either zero or $O(g) \sim 1$, as needed for radiative breaking. 

\subsection{Symmetry Breaking with an Extra $U(1)'$}
An additional non-anomalous $U(1)'$ gauge symmetry can be broken by the
VEV of a SM singlet $S$ with nonzero $U(1)'$ charge $Q_S$. The addition of the
$U(1)'$ and $S$ to the ordinary SM results in new arbitrary parameters in the
scalar potential, so there is in general no prediction for the $Z'$ mass scale.
However, things are much more constrained in the $U(1)'$ extension of
the MSSM~\cite{MSSMext}. Let us assume that $Q_1 + Q_2 \ne 0$, where $Q_{1,2}$
are the $U(1)'$ charges of $H_{1,2}$, so that $U(1)'$ invariance forces $\mu = 0$.
If $Q_1 + Q_2 +Q_S = 0$ one can have
\begin{equation}
W = h_S \hat{S} \hat{H}_1 \cdot \hat{H}_2 + h_Q \hat{u}^c_3 \hat{Q}_3 \cdot \hat{H}_2
+ \left[ h_D \hat{S} \hat{D}_1  \hat{D}_2 \right],
\label{MSSMUW}
\end{equation}
where the last term is an optional coupling of $S$ to new exotic multiplets $D_{1,2}$.
The analogue of (\ref{MSSMV}) becomes
\begin{eqnarray}
V_F & = & h_S^2 \left( |H_1|^2 |H_2|^2 + |S|^2 |H_1|^2 + |S|^2 |H_2|^2 \right) \nonumber \\ 
V_D & = & \frac{G^2}{8} \left( |H_1|^2 - |H_2|^2 \right)^2 +
\frac{g_1'^2}{2} \left( Q_1 |H_1|^2 + Q_2 |H_2|^2 +Q_S |S|^2 \right)^2 
\nonumber \\ 
V_S & = & m_1^2 |H_1|^2 + m_2^2 |H_2|^2 +m_S^2 |S|^2
- \left(A_S h_S S H_1 \cdot H_2 + {\rm h.c.} \right), 
\label{MSSMUV}
\end{eqnarray}
where $g_1'$ is the $U(1)'$ gauge coupling.
Thus, if $S$ acquires a VEV, one has an effective $\mu$ parameter
$\mu_{\rm eff} = h_S \langle S \rangle$, and the corresponding
$(B \mu)_{\rm eff} =A_S h_S \langle S \rangle$. Acceptable EW breaking
can  occur if $\langle S \rangle$ and $A_S$ are of $O$(TeV).
If all of the soft SUSY breaking parameters are of $O$(\mtt), then
one expects not only $\mu_{\rm eff}$ and $(B \mu)_{\rm eff}$ but also
$M_Z$ and $M_{Z'}$ to be of $O$(\mtt). Only some limiting (or 
somewhat tuned) cases will yield allowed $\theta$ and $M_{Z'}$.
The $Z-Z'$ mixing angle $\theta$ is given by
\begin{equation}
\theta=\frac{1}{2}\arctan\left(\frac{2\Delta^2}{M^{2}_{Z'}-M^{2}_{Z}}
\right),
\label{theta}
\end{equation}
where
\begin{eqnarray}
M_{Z}^{2}&=&\frac{1}{4}G^2(v_{1}^2+v_{2}^2),\\
M_{Z'}^{2}&=&{g}_{1}'^{2}(v_{1}^{2}Q_{1}^{2}+v_{2}^{2}Q_{2}^{2}+s^{2}Q_{S}^{2}),\\
\label{mix}
\Delta^{2}&=&\frac{1}{2}g_{1}'\,G(v_{1}^2Q_{1}-v_{2}^2Q_{2})
\end{eqnarray}
are respectively the $Z$ and $Z'$ mass squares in the absence of mixing
and the mixing mass squared, and $s \equiv \sqrt{2} \langle S \rangle$. 
Small mixing requires
small $\Delta$ and/or $M_Z \ll M_{Z'}$.

Two viable scenarios were described in \cite{cdeel}. (i) In
the {\it Large} $A_S$ {\it Scenario} the EW and $U(1)'$ breaking is
driven by a large $A_S h_S$ in the last term in (\ref{MSSMUV}).
This leads to $v_1 \sim v_2 \sim s$ (a generalization of $\tan \beta \sim 1$).
In the special case $Q_1 = Q_2 = -Q_S/2$ one finds $\theta \sim 0$
and the prediction $M_{Z'}^2/M_Z^2 \simeq 12 g_1'^2 Q_1/G^2$.
For example, a concrete model \cite{lw} with the couplings of the $E_6$ $\psi$
model \footnote{This model, which has the matter content and couplings of
three 27-plets as well as  two Higgs-like doublets from $ 27 + 27^\ast$, is anomaly
free and consistent with gauge unification. It is string-motivated, i.e.,
the Yukawa couplings are of $O(g)$ or zero, and the $B$ and $L$ violating
GUT Yukawa relations are not respected, so that $S$ can be light.}
yields $M_{Z'} \sim 84$ GeV. This model is not leptophobic \footnote{An
alternative $E_6$ model involving matter from an extra 78 has much larger kinetic
mixing and can lead to leptophobic couplings \cite{suppressed}.}
and is therefore
excluded, but it illustrates a scenario that may be viable in string-derived
models with suppressed couplings to ordinary fermions.

(ii) In the {\it Large} $S$ {\it Scenario} one assumes that all
of the soft parameters ($|m_{1,2,S}|, |A_S|$) are of $O$(1 TeV), with
$m_S^2 < 0$. Then $s^2 \sim -2 m_S^2/g_1'^2 Q_S^2$ and $M_{Z'}^2 \sim
-2 m_S^2$. One can have a smaller EW scale $v_{1,2} \ll s$ by accidental
cancellations (because of $V_D$ this often only involves one constraint).
To avoid excessive tuning this implies $M_{Z'} \simle O$(1 TeV).
Then $\theta \sim (\Delta^2/M_Z^2) (M_Z^2/M_{Z'}^2)$ is small due
to $M_Z^2 \ll M_{Z'}^2$, and can be further suppressed for small $\Delta^2$.

Both scenarios have a number of interesting consequences. These include:
(i) A solution to the $\mu$ problem \cite{u1mu}, with $\mu_{\rm eff}$
naturally of $O(M_Z)$ (large $A_S$) or $O$(TeV) (large $S$). (ii)
A $Z'$ and associated exotics with masses $\simle O$(TeV). (iii)
a predictive pattern of Higgs masses (large $A_S$) or weakened upper limit
on the lightest Higgs (large $S$) \cite{MSSMext}.
(iv) Characteristic shifts in the scalar masses due to the $U(1)'$
 $D$ term \cite{dshift}. (v) New dark matter possibilities (e.g., $\tilde{S}$)
\cite{ce}.

The weak scale parameters needed for both scenarios can be
generated by radiative breaking \cite{cdeel,keithma,lw}. As motivated by SUGRA,
we assume that at \mpl \ all of the scalar mass squares ($m_{1,2}^2, m_S^2,
m_{\tilde{l}}^2, m_{\tilde{q}}^2$) are positive and of $O$(\mtt$^2$),
but not necessarily universal. We also assume that the gaugino
masses $M_i$ and the $A$ parameters are of $O$(\mtt). The  coupled
one-loop RGE equations for the running gauge and Yukawa couplings and
the soft parameters $m^2, M, A$ were studied for various toy models \cite{cdeel}
and models with $E_6$ couplings \cite{lw} to relate the initial parameters at
\mpl \ to the EW scale parameters. It was found that the large $|A_S|$
scenario was possible though somewhat fine-tuned (it is necessary to ensure
moderate $|A_Q|$, the $A$ term associated with $h_Q$, to avoid dangerous
charge-color breaking minima). The large $S$ scenario, which requires
$m_S^2 < 0$ at the EW scale  is most easily obtained if there is
a Yukawa coupling of $S$ to exotic multiplets (the optional $h_D$ term
in  (\ref{MSSMUW})), but can  be obtained without such couplings for some
(non-universal) initial conditions.

\subsection{Intermediate Scale Breaking}
Another possibility is $U(1)'$ breaking associated with a
$D$-flat direction at an intermediate scale \cite{cceel1},
which is expected to occur in many string models and which
may be associated with fermion mass hierarchies\footnote{A
similar mechanism could occur for a total gauge singlet field.}. 
This can occur, for example, if there are two SM singlets
$S_{1,2}$ with $Q_{S_1} Q_{S_2} < 0$. If the model
is also $F$-flat at the renormalizable level (i.e., there
are no terms $\hat{S}_i \hat{S}_j$ or $\hat{S}_i \hat{S}_j
\hat{S}_k$ in $W$), the low energy potential for $S_{1,2}$  is
\begin{equation}
V(S_1,S_2) = m_1^2 |S_1^2| + m_2^2 |S_2^2| + \frac{g_1'^2}{2}
( Q_{S_1} |S_1^2| + Q_{S_2} |S_2^2|)^2,
\label{intpotential}
\end{equation}
where the quartic term vanishes for $|S_2^2|/|S_1^2|
= - Q_{S_1}/Q_{S_2}$. 

As an example, suppose 
$Q_{S_1}= -Q_{S_2}$, and further that at low energies $m_{S_1}^2 < 0$
and $m_{S_2}^2 > 0$, as would typically occur by the
radiative mechanism if $W$ contains a term 
$h_D \hat{S_1} \hat{D}_1  \hat{D}_2$. If $m^2 \equiv 
m_{S_1}^2 + m_{S_2}^2 >0$ the minimum will occur at
$\langle S_1 \rangle \ne 0, \ \langle S_2 \rangle = 0$.
Then, $\langle S_1 \rangle$
and $M_{Z'}$ will be at the EW scale ($\simle$ 1 TeV), just as
in the case of a single $S$. On the other hand, for $m^2 < 0$,
the potential along the $F$ and $D$ flat direction
$S_1 = S_2 \equiv S$ is
\begin{equation}
V(S) = m^2 S^2,
\label{unbounded}
\end{equation}
which appears to be unbounded from below. However,
$V(S)$ can be stabilized by either of two mechanisms \cite{cceel1}:
(a) The leading loop corrections to the effective (RGE-improved)
potential
result in $m^2 \rightarrow m^2(S)$ in (\ref{unbounded}). Since
$m^2$ runs from a positive value at \mpl \ to a negative value at
low energies, the RGE-improved potential will have a minimum
close to but slightly below the scale $\mu_{RAD}$ at
which $m^2$ goes through zero. It was shown in \cite{cceel1}
that $\mu_{RAD}$ can occur anywhere in the range $10^3 - 10^{17}$
GeV, depending on the soft breaking parameters and the exotic
($\hat{D}_i$) quantum numbers. (b) Another possibility is that 
the $F$-flatness is lifted by higher-dimensional
nonrenormalizable operators (NRO) in $W$, as are expected
in string models, such as $W = (\hat{S}_1 \hat{S}_2)^2/M$,
where $M \sim 10^{17}-10^{18}$ GeV is of the order of the
string scale. For example, if $W$ contains
\begin{equation}
W_S = \frac{\hat{S}^{K+3}}{M^K},
\label{flatnro}
\end{equation}
when evaluated along the flat direction $\hat{S}$, 
then the potential will be minimized at the scale
\begin{equation}
\mu_{NRO} \sim \left[ \mtt M^K \right]^{\frac{1}{K+1}},
\label{munro}
\end{equation}
which is around $10^{10}$ GeV for $K = 1$.

In general, both the radiative and NRO stabilization mechanisms
can occur, and $\langle S \rangle$ will be of the
order of the smaller of $\mu_{RAD}$ and $\mu_{NRO}$.
In both cases, one expects $M_{Z'}, M_{D_i} \sim \langle S \rangle$.
Also, $V''$ is of $O(\mtts)$ at the minimum, leading to
an electroweak scale invisible scalar. There are also
characteristic $D$-induced shifts in the effective soft masses
\cite{cceel1}. A effective $\mu$ parameter can be generated
by the superpotential term
\begin{equation}
W_\mu = \hat{S} \hat{H}_1 \cdot \hat{H}_2
\left( \frac{\hat{S}}{M} \right)^{P_\mu}
\Rightarrow \mu_{\rm eff} \sim \langle S \rangle
\left( \frac{\langle S \rangle}{M}
\right)^{P_\mu}.
\label{muntermro}
\end{equation}
(The special case $P_\mu=0$ is needed for the
EW scale breaking scenario.)
For radiative stabilization, one obtains the needed $\mu_{\rm eff} \sim$
1 TeV for, e.g.,  $P_\mu = 1$ and $\mu_{RAD} \sim 10^{10}$ GeV.
For NRO stabilization,
\begin{equation}
\mu_{\rm eff} \sim \mtt
\left(\frac{\mtt}{M}\right)^{\frac{ P_\mu-K}{K+1}},
\label{muvalue}
\end{equation}
which is of the order of the soft breaking (and EW) scale \mtt \
for $P_\mu = K$. 

Intermediate scale breaking scenarios have interesting implications
for quark, charged lepton, and neutrino masses \cite{cceel1,sterile}.
For example, a $u$-type quark mass may be generated by the term
\begin{equation}
W_u = h_u \hat{u}^c \hat{Q} \cdot \hat{H}_2 \left( \frac{\hat{S}}{M}
\right)^{P_u},
\label{wu}
\end{equation}
where in string models the nonzero coefficients $h_u$ are of $O(g) \sim 1$
for $P_u = 0$, and can be absorbed into $M$ for $P_u > 0$. (\ref{wu})
leads to an effective Yukawa coupling and fermion mass (in the case
of NRO stabilization)
\begin{equation}
y_u \sim \left( \frac{\langle S \rangle}{M}\right)^{P_u} \Rightarrow 
m_u \sim \left( \frac{\mtt}{M} \right)^{\frac{P_u}{K+1}}
\langle H_2 \rangle.
\label{yeff}
\end{equation}
Presumably, the $t$ mass is associated with $P_t = 0$ \cite{faraggi}, while the
$u$ and $c$ masses, and any inter-generational masses associated with
family mixing, could be due to operators of higher dimension. Similar
hierarchies of dimensions of operators could lead to small $d$ and $e$
type masses and mixings, especially for the first two families, as
well as naturally tiny Dirac neutrino masses, without the need
for invoking a seesaw \cite{cceel1}. Which terms actually have
non-zero coefficients is determined not only by gauge invariance in the
four dimensional effective field theory, but by string selection rules
as well \cite{cceel3}. This mechanism of small effective Yukawas
suppressed by intermediate scale VEV's is somewhat analogous to
the attempts \cite{anomalous,anommodel} to generate Yukawas suppressed by powers
of $\langle S_A \rangle/M \sim 1/10$, where $S_A$ is a field which
breaks the anomalous $U(1)'$ present in many free fermionic models \cite{free}.
However, the lower intermediate scale considered here allows for the use of lower
dimension operators\footnote{Most of the studies \cite{anommodel} have assumed that
the non-zero coefficients could be classified according to the anomalous
$U(1)_A$ symmetry. However, this is not the case in free fermionic
models \cite{cceel3}.}. It is also possible to generate Majorana masses $m_M$
for sterile ($SU(2)$-singlet) neutrinos $N^c_L$ by the operators
\begin{equation}
W_{\rm M} \sim \hat{N}^c_L \hat{N}^c_L \hat{S}
\left( {\hat{S}\over {M}} \right)^{P_M},
\label{singletmajorana}
\end{equation}
implying 
\begin{equation}
m_M  \sim  \langle S \rangle
\left( \frac{\langle S \rangle}{ M}
\right)^{P_M}   \sim
 \mtt
\left(\frac{\mtt}{M}\right)^{\frac{ P_M-K}{K+1}},
\label{majoranamass}
\end{equation}
which can be large (leading to a seesaw) or small, depending on
the sign of $P_M -K$. From (\ref{yeff}) and (\ref{majoranamass})
and the fact that $\langle H_{1,2} \rangle  \sim \mtt$ for
radiative breaking, one finds that neutrino Dirac and Majorana
masses can be naturally small and comparable in the special
case $P_D = P_M - K$, where $P_D$ is the power analogous to
$P_u$ in (\ref{wu}) for a Dirac neutrino mass term \cite{sterile}.
This can lead to significant mixing between ordinary neutrinos and
light sterile neutrinos, as is suggested phenomenologically
by the experimental hints of neutrino mass \cite{hints}.

\section{Perturbative String Vacua}
The work discussed in Section \ref{stringmotivated} was motivated
by certain general features of perturbative string models, especially
(a) the existence of additional $U(1)$'s and exotics, (b) that the Yukawa 
couplings at the string scale are either zero or of $O(g) \sim 1$, (c)
that world-sheet selection rules often forbid terms in the superpotential $W$
that would be allowed by the  gauge symmetries of the effective four-dimensional
field theory, and
(d) that there are no elementary bilinear (mass) terms in $W$. A more ambitious
project is to try to derive the consequences of  specific string vacua. There are
many possible string vacua, and none that have been studied are fully
realistic. However, there are models based on the free fermionic
construction \cite{models,free} that are quasi-realistic, containing the
ingredients of the MSSM (gauge group, and candidates for three 
ordinary families and two Higgs doublets) and some form of gauge unification.
They typically also contain a (partially) hidden sector non-abelian group,
an anomalous $U(1)'$, a number of extra non-anomalous $U(1)'$s, and many additional
exotic chiral supermultiplets. The latter include non-chiral exotic multiplets,
fractionally charged states, and mixed states transforming non-trivially under
both the ordinary and hidden sector groups.

A first step in studying the low energy consequences of such models is to
determine which fields acquire VEVs at or near the string scale, in a way 
that breaks the anomalous $U(1)'$ but maintains $D$ and $F$ flatness.
Techniques have recently been developed to compute classes of $D$-flat
directions that can be proved $F$-flat to all orders \cite{cceel3}.
A number of models were considered, and it was found that those which
have such flat directions composed on non-abelian singlet fields typically
leave one or more non-anomalous $U(1)'$s unbroken at the string scale.
A next step, currently in progress \cite{everett}, is to study the effective
superpotential of the resulting model in these flat directions, after replacing the
scalar fields which appear in the flat directions by their VEVs.
In particular, it will be possible to study the $U(1)'$ breaking patterns
and low energy consequences, after making appropiate ans\"{a}tze for
the soft supersymmery breaking terms and the structure of the K\"{a}hler
potential. It is unlikely that any realistic models will be found, but it is
hoped that the analysis will give useful insights into the type of
physics consequences that may derive from perturbative string theories.

\section*{Acknowledgments} 
It is a pleasure to thank J. Cleaver, M. Cveti\v c, D. Demir,
J. R. Espinosa, L. Everett, and J. Wang for fruitful and enjoyable collaborations.
This work was supported by U.S. Department of Energy Grant No. DOE-EY-76-02-3071.

\end{document}